\documentclass[11pt]{article}
\usepackage{moriond,epsfig}
\def\Journal#1#2#3#4{{#1} {\bf #2}, #3 (#4)}

\def\NIMA{{\em Nucl. Instrum. Methods} A}

\def\NPPS{{\em Nucl. Phys. Proc. Suppl.}}
\def\PLB{{\em Phys. Lett.}  B}
\def\PRL{\em Phys. Rev. Lett.}
\def\PRD{{\em Phys. Rev.} D}

\RequirePackage{xspace}

\def\pep2{PEP-II}
\def\Bbar    {\kern 0.18em\overline{\kern -0.18em B}{}\xspace}
\def\Bz      {\ensuremath{B^0}\xspace}
\def\Bzb     {\ensuremath{\Bbar^0}\xspace}
\def\Bpm     {\ensuremath{B^\pm}\xspace}
\def\Bu      {\ensuremath{B^+}\xspace}
\def\Bub     {\ensuremath{B^-}\xspace}
\def\Bp      {\ensuremath{\Bu}\xspace}
\def\Bm      {\ensuremath{\Bub}\xspace}
\def\BB      {\ensuremath{B\Bbar}\xspace} 

\def\Dbar    {\kern 0.2em\overline{\kern -0.2em D}{}\xspace}
\def\Dz      {\ensuremath{D^0}\xspace}
\def\Dzb     {\ensuremath{\Dbar^0}\xspace}

\def\etal{{\it et~al.,}}

\def\be{\begin{equation}}
\def\ee{\end{equation}}
\def\bea{\begin{eqnarray}}
\def\eea{\end{eqnarray}}

\begin{document}
\vspace*{4cm}
\title{PROSPECTS FOR MEASURING GAMMA AT BABAR}

\author{ F.F. Wilson \\ representing the BaBar Collaboration}

\address{H H Wills Physics Lab., Tyndall Avenue, Bristol, BS8 1TL, UK}

\maketitle\abstracts{The prospects for measuring the angle $\gamma$
with present day $B$ factories are examined. A number of approaches
are discussed with reference to recent preliminary measurements 
based on a data sample of approximately 88 million \BB\ pairs collected by the BaBar
detector at the \pep2\ asymmetric $B$ Factory at SLAC.}

\section{Introduction}

Until recently it was thought that the measurement of the angle
$\gamma$ of the Cabibbo-Kobayashi-Maskawa (CKM) matrix was beyond the
reach of present day $B$ factories such as BaBar and Belle. A
number of experimental and theoretical developments have resulted in a
change in attitude. The speed at which the
angles $\beta$ and $\alpha_{eff}$ have been measured has confirmed
that the current experiments can extract the
maximum information from the data. At the same
time, both Belle and BaBar are aiming to exceed their design luminosities
by considerable amounts; BaBar expects to take 500 fb$^{-1}$ by 2006 and
1 at$^{-1}$ by the time LHC turns on.
On the theoretical side, advances have been made in factorisation and
flavour--symmetry models. Branching fraction (BF) measurements at
$B$ factories have added extra constraints to theses models. At the
same time, our understanding of electro-weak penguins and rescattering
have given us greater confidence in our ability to directly confront
the predictions of the Standard Model \cite{rosner}.

BaBar has reported measurements of $\sin(2\beta)$ and
$\sin(2\alpha_{eff})$ where $\mid\alpha-\alpha_{eff}\mid$ is expected to be less than
51$^{\circ}$ using the conservative Grossman-Quinn bound
\cite{angles,quinn}. From these constraints, the range $50^{\circ} < \gamma
< 70^{\circ}$ would appear to be favoured. But to really understand this angle, we
must make a number of direct measurements.

In the following sections, a number of possible methods are
discussed. Although only preliminary results from BaBar are reported,
Belle has a similar physics reach.

\section{Preliminary Measurements at BaBar}

A description of the BaBar detector is given here \cite{babar}.
All the following measurements contain some or more of the following elements. Beam
constraints are used to define a signal region. Background from
continuum events are suppressed by using a series of event shape
variables often in the form of Fisher discriminants or neural
nets. Particle Identification is performed using energy loss in the
tracking detectors and the calorimeter and the Cherenkov angle in the
DIRC. If required, the two B decay vertices are reconstructed and
flavour tagging performed. Finally a global maximum likelihood
technique is used to achieve the greatest sensitivity. 

\subsection{$\Bpm \rightarrow D_{CP}K^{\pm}$}
\label{subsec:dcp}

\noindent The CP eigenstates $\mid D_{\pm}^0\rangle$ of the neutral D
meson system with CP eigenvalues $\pm 1$ are given by:

\begin{equation}
  \mid D_{\pm}^0\rangle = \frac{1}{\sqrt{2}} \left( \mid D^0\rangle
  \pm \mid \Dzb \rangle \right)
\end{equation}

so that the $\Bpm \rightarrow D^0_+K^\pm$ transition amplitudes can
be expressed as:

\begin{eqnarray}
\sqrt{2} A(\Bp \rightarrow D^0_+K^+) = A (\Bp \rightarrow D^0 K^+) +
A (\Bp \rightarrow \Dzb K^+) \\ \nonumber
\sqrt{2} A(\Bm \rightarrow D^0_+K^-) = A (\Bm \rightarrow \Dz K^-) +
A (\Bm \rightarrow \Dzb K^-)
\end{eqnarray}

These relations are exact, originate from pure tree decays and receive
no contributions from penguins.  They can be represented by 2
triangles in the complex plane. Since the transition amplitude A($\Bp
\rightarrow \Dzb K^+)$ = A($\Bm \rightarrow D^0K^-)$ and the
difference in CP-violating weak phase between the $\Bp \rightarrow
D^0K^+$ and the $\Bm \rightarrow \Dzb K^-$ amplitudes is proportional
to $e^{2i\gamma}$, these triangles allow a determination of $\gamma$
by measuring the six amplitudes.  A complementary method uses $\Bz
\rightarrow D^0_+K^{*0}$, $\Bz \rightarrow \Dzb K^{*0}$ and $\Bz
\rightarrow D^0K^{*0}$.

BaBar has measured the ratio of the branching fractions $\Bm
\rightarrow D^0K^-$ and $\Bm \rightarrow D^0\pi^-$ to be
$8.31\pm0.35\pm0.2\%$ with the individual ratios for the sub--decays 
$\Dz \rightarrow K^-\pi^+$,
$\Dz \rightarrow K^-\pi^+\pi^+\pi^-$ and $\Dz \rightarrow K^-\pi^+\pi^0$
to be $8.4\pm0.5\pm0.2\%$, $8.7\pm0.7\pm0.3\%$ and $7.7\pm0.7\pm0.2\%$
respectively. The ratio of the branching fractions for the decay
$\Bpm \rightarrow D^0_+K^{\pm}$ to $\Bpm \rightarrow D^0_+\pi^{\pm}$ is
$7.42\pm 1.7\pm 0.6\%$ for $D^0_+\rightarrow K^-K^+$ and
$12.9\pm4.0^{+1.1}_{-1.5}\%$ for $D^0_+\rightarrow \pi^-\pi^+$. The
weighted average is $8.8\pm1.6\pm0.5\%$. The CP asymmetry is measured
to be \cite{Dcp}:

\begin{equation}
  A_{CP} = 
\frac{BF(B^-\rightarrow D^0_+K^-) - BF(B^+\rightarrow D^0_+K^+) }
     {BF(B^-\rightarrow D^0_+K^-) + BF(B^+\rightarrow D^0_+K^+) } =
0.07\pm 0.17 \pm 0.06
\end{equation}

\subsection{$B \rightarrow \pi\pi, \pi K$}
\label{subsec:pipi}

\noindent An alternative method is to use SU(3) flavour symmetries to
derive amplitude relationships between $B$ decays into $\pi\pi$, $\pi K$
and $K\overline{K}$. The amplitudes for the decays $B^+ \rightarrow
\pi^+\pi^0$, $B^+ \rightarrow \pi^+ K^0$ and  $B^+ \rightarrow K^+
\pi^0$ can be expanded in terms of colour-allowed and colour-suppressed
tree diagrams and QCD penguin diagrams for both strangeness--preserving
and strangeness--changing reactions. Neglecting electroweak penguins,
this can be written in terms two triangles again:

\begin{eqnarray}
\sqrt{2} A(B^+ \rightarrow \pi^0 K^+) + A (B^+ \rightarrow \pi^+ K^0) =
r_u \sqrt{2} A (B^+ \rightarrow \pi^0 \pi^+) \\
\sqrt{2} A(B^- \rightarrow \pi^0 K^-) + A (B^- \rightarrow \pi^- K^0) =
r_u \sqrt{2} A (B^- \rightarrow \pi^0 \pi^-)
\end{eqnarray}

where $r_u$ is approximately the ratio of strangeness-changing to
strangeness--preserving tree diagrams. Since no tagging or
time--dependent analyses are required this is a promising experimental
measurement.

The six sides of the two triangles have been measured by BaBar. Table
\ref{tab1} shows the 90\% confidence upper limits, branching fractions
and CP asymmetries for a number of $\pi K$ modes
\cite{twobody}. Constraints on the angle $\gamma$ can be derived from
the ratio of the branching fractions of a number of these modes
\cite{ratios}.

\begin{table} 
\begin{center} 
\begin{tabular}{|l|l|l|} \hline
Mode & Branching Fraction ($\times 10^{-6}$) & A$_{\mbox{CP}}$ \\ \hline
$\Bz \rightarrow K^+\pi^-$ & 
$17.9 \pm 0.9 \pm 0.7$ &
$-0.102 \pm 0.050 \pm 0.016$ \\ 
$\Bp \rightarrow \pi^+\pi^0$ & 
$5.5^{+1.0}_{-0.9} \pm 0.6$ &
$-0.03^{+0.18}_{-0.17} \pm 0.02$ \\ 
$\Bp \rightarrow K^+\pi^0$ & 
$12.8^{+1.2}_{-1.1} \pm 1.0$ &
$-0.09 \pm 0.09 \pm 0.01$ \\ 
$\Bp \rightarrow K^0\pi^+$ & 
$17.5 \pm 1.8 \pm 1.3$ &
$-0.17 \pm 0.10 \pm 0.02$ \\ 
$\Bz \rightarrow K^0\pi^0$ & 
$10.4 \pm 1.5 \pm 0.8$ &
$0.03 \pm 0.36 \pm 0.09$ \\ 
$\Bz \rightarrow \pi^+\pi^-$ & 
$4.6 \pm 0.6 \pm 0.2$ & \\ 
$\Bz \rightarrow \pi^0\pi^0$ & 
$<3.6$ & \\ 
$\Bz \rightarrow K^+K^-$ & 
$<0.6$ & \\ 
$\Bz \rightarrow K^+K^0$ & 
$<1.3$ & \\ 
\hline
\end{tabular} 
\caption{\label{tab1}Branching Fractions,  CP asymmetries and 90\% confidence
upper limits for $B\rightarrow K\pi$, $\pi\pi$ and $KK$ decays.}
\end{center} 
\end{table}

\subsection{$\sin(2\beta+\gamma)$ from $B^0\rightarrow D^{(*)-}\pi^+$}
\label{subsec:sin2bg}

\noindent If the value of $\sin(2\beta)$ is taken from other
measurements then $\gamma$ can be extracted from measurements of
$\sin(2\beta+\gamma)$ in the decays of $\Bz \rightarrow
D^{(*)\pm}\pi^{\mp}$. The CP asymmetries are expected to be small but
the decay rates are large. Both $D^{(*)+}\pi^-$ and $D^{(*)-}\pi^+$
are accessible from \Bz\ and \Bzb\ decays. The angle $\gamma$ enters
through the Cabibbo--suppressed amplitude V$_{\mbox{ub}}$. The
branching fraction $\Bz \rightarrow D^{(*)+}\pi^-$ has not been
measured but it can be estimated from the measurement of $\Bz
\rightarrow D^{(*)+}_s\pi^-$ and the SU(3) symmetry relation \cite{dun}:

\begin{equation}
   BF(\Bz \rightarrow D_s^{(*)+}\pi^-) \equiv \frac{BF(\Bz \rightarrow
   D^{(*)-}\pi^+)}{\tan^2\theta_c} \frac{f^2_{D_s^{(*)}}} {f^2_{D^{(*)}}} \lambda^2_{D^{(*)}\pi}
\end{equation}
where $\theta_c$ is the Cabibbo angle, $f$ are calculated form factors \cite{bec} and
$\lambda_{D^{(*)}\pi}$ is given by:

\begin{equation}
  \lambda_{D^{(*)}\pi} =
\frac{A(\Bzb \rightarrow D^{(*)+}\pi^-)} 
     {A(\Bz  \rightarrow D^{(*)+}\pi^-)} e^{-2\beta}
= \mid \lambda_{D^{(*)}\pi} \mid e^{-i(2\beta+\gamma)}
\end{equation}

$\lambda_{D\pi}$ is measured by BaBar to be $0.021^{+0.004}_{-0.005}$ and
$\lambda_{D^*\pi}= 0.017^{+0.005}_{-0.007}$. An additional error of
30\% is required for uncertainties in the SU(3) symmetry breaking. In
addition the following branching fractions and 90\% confidence upper
limits have been measured \cite{sin2beta}: 
$B^0\rightarrow D^+_s\pi^- = 3.1\pm1.0\pm1.0 \times 10^{-5}$,
$B^0\rightarrow D^{*+}_s\pi^- < 4.1 \times 10^{-5}$,
$B^0\rightarrow D^-_s K^+ = 3.2\pm1.0\pm1.0 \times 10^{-5}$ and
$B^0\rightarrow D^{*-}_s K^+ < 2.5 \times 10^{-5}$.

\subsection{$B^{\pm} \rightarrow \pi\pi\pi, \pi \pi K$}
\label{subsec:pipipi}

\noindent It has been proposed that Dalitz plots in charmless 3-body
decays can be used to extract $\gamma$ \cite{bajc}. 
The Dalitz plot for $B^\pm\rightarrow
\pi^+\pi^-\pi^\pm$ has a number of resonances with amplitudes
proportional to $V_{\mbox{ub}}V_{\mbox{ud}} \sim e^{-i\gamma}$. Providing there
are at least 2 weak and 2 strong phases, the angle $\gamma$ can be
extracted from interference between the resonances and
non--resonance decays. There are some approximations concerning the relative
contributions of tree and penguin diagrams; this can be compensated for
by using the $B^\pm\rightarrow K^+\pi^-\pi^\pm$ Dalitz plot and SU(3) symmetry relations.

BaBar has measured the inclusive rates and CP asymmetries for $B^\pm\rightarrow
h^+h^-h^\pm$ where h=$\pi$,K and exclusive rates for $B^\pm\rightarrow
K^+\pi^-\pi^\pm$. These results are tabulated in Table \ref{tab2} \cite{threebody}.

\begin{table} 
\begin{center} 
\begin{tabular}{|l|l|l|} \hline
Mode & Branching Fraction ($\times 10^{-6}$) & A$_{\mbox{CP}}$ \\ \hline
$\Bp \rightarrow \pi^+\pi^-\pi^+$ & 
$10.9 \pm 3.3 \pm 1.6$ &
$-.039 \pm 0.33 \pm 0.12$ \\ 
$\Bp \rightarrow K^+\pi^-\pi^+$ & 
$59.2 \pm 3.8 \pm 3.2$ &
$0.01 \pm 0.07 \pm 0.03$ \\ 
$\Bp \rightarrow K^+ K^- K^+$ & 
$29.6 \pm 2.1 \pm 1.6$ &
$0.02 \pm 0.07 \pm 0.03$ \\ 
$\Bp \rightarrow K^+ K^- K^+$ & 
$< 6.3$ & \\ 
$\Bp \rightarrow \pi^+ K^- \pi^+$ & 
$< 1.8$ & \\ 
$\Bp \rightarrow K^+ \pi^- K^+$ & 
$< 1.3$ & \\ 
\hline
$\Bp \rightarrow K^{*0}(892)\pi^+, K^{*0} \rightarrow K^-\pi^-$  & 
$10.3\pm 1.2^{+1.0}_{-2.7}$ & \\
$\Bp \rightarrow f_{0}(980)K^+, f_{0} \rightarrow \pi^+\pi^-$  & 
$9.2\pm 1.2^{+2.1}_{-2.0}$ & \\
$\Bp \rightarrow \chi_{c0} K^+, \chi_{c0} \rightarrow \pi^+\pi^-$  & 
$1.46\pm 0.35 \pm 0.12$ & \\
$\Bp \rightarrow \Dz \pi^+, \Dz \rightarrow K^+\pi^-$  & 
$184.6\pm 3.2 \pm 9.7$ & \\
$\Bp \rightarrow \mbox{higher}\ K^{*0}\pi^+, K^{*0} \rightarrow K^+\pi^-$  & 
$25.1\pm 2.0 ^{+11.0}_{-5.7}$ & \\
$\Bp \rightarrow \rho_0(770) K^+, \rho_0 \rightarrow \pi^+\pi^-$  & 
$<6.2$& \\
$\Bp \rightarrow K^+\pi^+\pi^-\ \mbox{(non--resonant)}$  & 
$<17$& \\
$\Bp \rightarrow \mbox{higher f}\ K^+, f \rightarrow \pi^+\pi^-$  & 
$<12$& \\ \hline
\end{tabular} 
\caption{\label{tab2}Branching Fractions,  CP asymmetries and 90\% confidence
upper limits for inclusive 3-body decays to $h^+h^-h^+$ ($h=\pi,K$)
and exclusive decays to $K^+\pi^-\pi^+$.}
\end{center} 
\end{table}

\section{Conclusion}

Progress has been made on a number of approaches to the measurement of
the angle $\gamma$. The ultimate accuracy is hard to
estimate as the error on
$\gamma$ depends on the value of $\gamma$ but meaningful measurements
of the angle $\gamma$ should be possible by the end of the lifetime of
the current experiments.

\section*{Acknowledgements}

I wish acknowledge the help I received from my BaBar colleagues in
preparing this talk, the international funding agencies for supporting
this research and the organisers and secretariat for another
successful conference.

%We are grateful for the extraordinary contributions of our \pep2\
%%colleagues in achieving the excellent luminosity and machine
%conditions that have made this work possible.  The success of this
%project also relies critically on the expertise and dedication of the
%computing organisations that support BaBar.  The collaborating
%institutions wish to thank SLAC for its support and the kind
%hospitality extended to them.  This work is supported by the US
%Department of Energy and National Science Foundation, the Natural
%Sciences and Engineering Research Council (Canada), Institute of High
%Energy Physics (China), the Commissariat \`a l'Energie Atomique and
%Institut National de Physique Nucl\'eaire et de Physique des
%Particules (France), the Bundesministerium f\"ur Bildung und Forschung
%and Deutsche Forschungsgemeinschaft (Germany), the Istituto Nazionale
%di Fisica Nucleare (Italy), the Foundation for Fundamental Research on
%Matter (The Netherlands), the Research Council of Norway, the Ministry
%of Science and Technology of the Russian Federation, and the Particle
%Physics and Astronomy Research Council (United Kingdom).  Individuals
%have received support from the A. P. Sloan Foundation, the Research
%Corporation, and the Alexander von Humboldt Foundation.

%\section*{Appendix}

\end{document}